\documentclass[doublespacing]{elsart}
\usepackage{amsmath,amssymb,latexsym}
\usepackage{graphics}
\RequirePackage{keyval, hyperref}
\hypersetup{hyperfigures=true,colorlinks=true,citecolor=blue,filecolor=red}

\newcommand{\transpose}[1]{#1^{\sf T}}
\newcommand{\abs}[1]{\left\vert#1\right\vert}
\newcommand{\avg}[1]{\langle#1\rangle}
\newcommand{\seq}[1]{\{#1\}}
\newcommand{\set}[1]{\{#1\}}

\renewcommand{\d}{\mathrm{d}}
\renewcommand{\e}{\mathrm{e}}

\begin{document}

\begin{frontmatter}

\title{Derivation of the Onsager Principle from Large Deviation Theory}

\author{Brian R. La Cour\corauthref{LaCour}}
\corauth[LaCour]{Corresponding author.}
\ead{blacour@arlut.utexas.edu}
\address{Applied Research Laboratories, The University of Texas at
Austin, \\ Austin, Texas 78758}
\author{William C. Schieve}
\address{Ilya Prigogine Center for Studies in Statistical Mechanics
and Complex Systems, Department of Physics, The University of Texas at
Austin, Austin, Texas 78712}

\begin{abstract}
The Onsager linear relations between macroscopic flows and
thermodynamics forces are derived from the point of view of large
deviation theory.  For a given set of macroscopic variables, we
consider the short-time evolution of near-equilibrium fluctuations,
represented as the limit of finite-size conditional expectations.  The
resulting asymptotic conditional expectation is taken to represent the
typical macrostate of the system and is used in place of the usual
time-averaged macrostate of traditional approaches.  By expanding in
the short-time, near-equilibrium limit and equating the large
deviation rate function with the thermodynamic entropy, a linear
relation is obtained between the time rate of change of the macrostate
and the conjugate initial macrostate.  A Green-Kubo formula for the
Onsager matrix is derived and shown to be positive semi-definite,
while the Onsager reciprocity relations readily follow from time
reversal invariance.  Although the initial tendency of a macroscopic
variable is to evolve towards equilibrium, we find that this evolution
need not be monotonic.  The example of an ideal Knundsen gas is
considered as an illustration.
\end{abstract}


\date{5 May 2003}

\end{frontmatter}


\section{Introduction}
\label{sec:intro}

Nonequilibrium thermodynamics traditionally studies both the
steady-state flow of a system maintained out of equilibrium, e.g., by
a temperature gradient, and the detailed approach to equilibrium of
transient fluctuations
\cite{Chapman_and_Cowling,deGroot,deGroot_and_Mazur}.  In this paper
we focus on the latter phenomenon, that is, the time evolution of
small fluctuations from a certain equilibrium state, defined as the
long-time expected or \textit{a priori} most-likely value of a given
set of macroscopic variables.  One way of quantifying this transient
behavior is through the determination of transport coefficients, which
may be used to characterize the rate of approach to equilibrium.

A well-known example of such transport coefficients is provided by the
linear theory of Onsager \cite{Onsager1931a,Onsager1931b} relating
macroscopic flows to thermodynamic forces.  This relation, known as
the Onsager principle, has become a cornerstone of nonequilibrium
statistical mechanics and has seen wide application in a diverse range
of fields.  The purpose of this paper, however, is to establish
Onsager's principle on a rigorous basis and in a fairly general
context.  In doing so, we hope to understand more clearly its meaning
and range of validity.

The key insight to our approach is the introduction of a generalized
free energy function which may be used to define the time-dependent
macroscopic variables of interest.  This approach has been applied
successfully to the broader problem of macroscopic determinism in
interacting and noninteracting systems
\cite{LaCour_and_Schieve2002,LaCour_and_Schieve2000} and relies
chiefly on rigorous results from the mathematical theory of large
deviations
\cite{LaCour_and_Schieve2002,Dembo_and_Zeitouni,Deuschel_and_Stroock,Ellis}.
Under suitable regularity conditions, one may construct a mapping,
$\psi_t$, from an initial macrostate $a_0$ at time zero to a predicted
macrostate $\psi_t(a_0)$ at time $t$.  It may be shown that the actual
macroscopic observable, $G_{n,t}$, considered as a phase function for
a system with finitely-many degrees of freedom, converges in
probability to the value $\psi_t(a_0)$ when conditioned on the initial
value $a_0$.

The approach used here is simply to expand $\psi_t(a_0)$ in a Taylor
series about the equilibrium macrostate and initial time.  Assuming
the function is smooth enough that this may be done, identification of
the equilibrium large deviation rate function with the thermodynamic
entropy allows us to derive, for short times, the linear Onsager
relation between macroscopic flows and thermodynamics forces.

Section~\ref{sec:NET} begins with a brief overview of the Onsager
principle in the context of nonequilibrium thermodynamics and
discusses prior work in its statistical formulation and derivation.
Section~\ref{sec:OLR} contains the main results of this paper, which
is a derivation of the Onsager principle, valid in the short-time,
near-equilibrium regime, using techniques from large deviation theory.
In Sec.~\ref{sec:GKF} a Green-Kubo formula for the Onsager matrix in
terms of the covariance matrix of the macroscopic variables is derived
and, in the case of time reversal invariance, shown to be symmetric.
The forward Onsager matrix is shown to be positive semidefinite as a
consequence of measure preservation, and the positivity of the entropy
production rate is discussed in relation to the second law of
thermodynamics.  We end in Sec.~\ref{sec:IGE} with the example of an
ideal gas undergoing specular reflection in a closed box.  The Onsager
coefficient for the fraction of particles on one side of a uniform
partition is computed in terms of the \textit{a priori} velocity
distribution.  Using Maxwellian and uniform velocity distributions,
respectively, the time-dependent entropy is computed numerically and
the results discussed.  Concluding remarks may be found in
Sec.~\ref{sec:Discussion}


\section{Nonequilibrium Thermodynamics}
\label{sec:NET}

Before addressing the problem of transport properties, we must first
establish a connection between the various mathematical objects we
have considered thus far and the more familiar objects of
nonequilibrium thermodynamics.  The traditional approach takes
concepts and relations from equilibrium thermodynamics and assumes a
range of validity into the nonequilibrium regime.  This extension is
largely justified by empirical success, and generally works best when
the system is near equilibrium.  Far from equilibrium, well-defined
dissipative structures may form and evolve in time
\cite{Prigogine1980,Prigogine1967}, but such structures are not the
object of study here.

Consider a system for which we are interested in the time evolution of
one or more macroscopic observables.  Let $\alpha(t)$ be the
vector-valued observable at time $t$.  (Later, we shall identify
$\alpha(t)$ with the asymptotically most probable macrostate at time
$t$.)  At equilibrium the entropy, $S$, is of course a well-defined
function of the given macroscopic variables.  By replacing the
equilibrium macrostates with $\alpha(t)$, we obtain a time-dependent
entropy $S(\alpha(t))$. If $S$ and $\alpha$ are differentiable, then
from the chain rule we have the well-known formula
\begin{equation}
\dot{S}(\alpha(t)) = X(\alpha(t)) \, J(t),
\label{eqn:entropy_rate}
\end{equation}
where $X = \transpose{\nabla}S$ is the so-called \emph{thermodynamic
force} and $J = \dot{\alpha}$ is the time rate of change or
\emph{flow} of the observable.  (We use $\nabla$ to denote a column
vector gradient and $\transpose{\nabla}$ for a row vector gradient.)
In what follows, $J(t)$ will be considered a $d$-dimensional column
vector, while $X(\alpha(t))$ is considered a row vector.

With the aforementioned differentiability assumptions, Eq.\
(\ref{eqn:entropy_rate}) is nothing more than a definition of $X$ and
$J$.  In the absence of phase transitions, it is indeed reasonable to
suppose $S$ is a differentiable function of the macrostates.  The
differentiability of $\alpha$, however, is less clear, and this
difficulty stems largely from the unclear and imprecise meaning of
$\alpha(t)$ itself.

An obvious candidate is for $\alpha(t)$ is the actual, time-varying
observable, $G_{n,t}$.  In this case, $J(t)$ is the instantaneous time
derivative, if it exists, of a particular realization of the
corresponding stochastic process.  However, such detailed time
variation surely cannot be part of a general macroscopic law.

In his 1952 paper, M. Green \cite{Green1952} considered using the
ensemble average, $\avg{G_{n,t}}_{a_0}$, of $G_{n,t}$ conditioned on
$G_{n,0} = a_0$ as a macroscopic variable, implicitly identifying this
with its large-$n$ probabalistic limit.  For macroscopic flows,
Casimir \cite{Casimir1945} and, later, Mori \cite{Mori1958} suggested
that $J(t)$ be defined as the time average rate of change of this
ensemble average over a mesoscopic time $\tau$, i.e., one intermediate
between the time between collisions and the relaxation time of the
system as a whole.  More specifically, they proposed that $J(t)$ be
defined as
\begin{equation}
J(t) = \frac{\avg{G_{n,t+\tau}}_{a_0} - \avg{G_{n,t}}_{a_0}}{\tau},
\end{equation}
where $\avg{\,\cdot\,}_{a_0}$ is an expectation with respect to the
initial conditional distribution for which $G_{n,t_0} = a_0$.  The
difficulty with such a proposal, however, is that it requires an
explicit time scale be incorporated into the basic definition of
$J(t)$.

Since what is desired of $J(t)$ is a measure of the typical,
macroscopic rate of change of the observables, a reasonable and more
elegant proposal is to equate $\alpha(t)$ with $\psi_t(a_0)$, the
large-$n$ expected value of the observable.  This has the twin virtue
of remaining faithful to the actual dynamics, since we assume
$G_{n,t}$ converges to $\psi_t(a_0)$ in probability, while remaining
insensitive to detailed time variations by taking, in effect, an
ensemble average.  In addition, this choice is free of the need to
specify an ad hoc time scale for $J(t)$.  Of course, for $J(t)$ to be
well defined we must still have that $\psi_t(a_0)$ is differentiable
in $t$, but that is a far more reasonable assumption.  (Later we shall
see that the one-sided time derivatives at the initial time, $t=0$,
will usually not agree, so care must be taken here to distinguish
between forward and reverse time derivatives.)

In 1931, Lars Onsager \cite{Onsager1931a,Onsager1931b} proposed a
simple, linear relationship between flows and forces based on a
generalization of well-known phenomenological laws such as the
thermoelectric effect studied earlier by Thompson.  Onsager postulated
a relationship of the form
\begin{equation}
J(t) = L \, \transpose{X(\alpha(t))},
\label{eqn:Onsager_relation}
\end{equation}
where $L$ is known as the \emph{Onsager matrix}.  Using time reversal
invariance (or ``microscopic reversibility,'' in his terminology),
Onsager showed that $L$ is symmetric for observables which are even
functions of the momentum, a result known as the Onsager reciprocity
relation.  The papers by Casimir \cite{Casimir1945} and Wigner
\cite{Wigner1954} give a detailed discussion of this latter
derivation.

Elements of the Onsager matrix may be identified with various
transport coefficients associated with the macroscopic variables of
interest.  These, in turn, may be derived in terms of the correlations
between macrostates at different times.  The term ``Green-Kubo
formula'' is here refers to any formula giving elements of the Onsager
matrix in terms of time-correlations of observables, owing to the
early work of Green \cite{Green1950} and Kubo \cite{Kubo1957} in
deriving such relations.

Recently, Oono \cite{Oono1993} has reconsidered the Casimir-Mori
proposal in the context of large deviation theory to obtain the
Onsager relation of Eq.~(\ref{eqn:Onsager_relation}).  Like Mori,
Oono considers a macroscopic time evolution which is coarse-grained in
time according to the postulated mesoscopic time scale $\tau$.  The
approach he takes is to write the macroscopic variable $\alpha(t)$ as
a time average over the interval $[t,t+\tau]$ of the observable
denoted here by $G_{n,t}$.  (The macroscopic flow is defined
similarly.)  A large deviation principle for the empirical time
average (valid for $\tau$ large but $n$ fixed) is then assumed, from
which an LDP for the macroscopic flow is obtained via the contraction
principle \cite{Dembo_and_Zeitouni}.  To get the right initial
distribution, Oono appeals to information theory to argue that the
most likely macroscopic flow should be in the form a canonical
expectation in which, following Mori, the Lagrange multipliers are
determined by the given initial flow.  (By contrast, this canonical
form was derived in \cite{LaCour_and_Schieve2002} using a conditional
LDP theorem \cite{LaCour_and_Schieve2002}.)  The connection to the
traditional thermodynamic quantities is then made in the usual way by
equating the large deviation rate function with the thermodynamic
entropy using the Einstein fluctuation formula.  The result which Oono
obtains is
\begin{equation}
\begin{split}
L &= \int_0^{\tau} \avg{\dot{G}_{n,s}\transpose{\dot{G}_{n,0}}} \d s \\
  &= \avg{[G_{n,\tau}-G_{n,0}]\dot{G}_{n,0}} \xrightarrow{\tau\to\infty}
     -\avg{G_{n,0}\transpose{\dot{G}_{n,0}}},
\end{split}
\label{eqn:Oono}
\end{equation}
assuming $\avg{G_{n,0}} = 0$ and the correlation
$\avg{G_{n,\tau}\dot{G}_{n,0}}$ vanishes for $\tau$ large.

Large deviation theory has also been used recently by Bertini et al.\
\cite{Bertini2002} to describe macroscopic fluctuations in the density
field of a Markovian lattice.  In the appropriate scaling limit,
fluctuations of the asymptotically most likely density field,
analogous to our $\psi_t(a_0)$, are shown to satisfy the
Onsager-Machlup principle; i.e., the fluctuations are time symmetric.
The present work differs in that we consider a general, deterministic
microscopic dynamics rather than an underlying Markovian microstate.
Our approach, however, will accomodate a fundamentally stochastic
system as well as it will a deterministic one.

Finally, we note that Gallavotti \cite{Gallavotti1996} has recently
derived the Onsager relations and Green Kubo formula for thermostated
systems as a consequence of the fluctuation
theorem\cite{Gallavotti_and_Cohen1995a,Gallavotti_and_Cohen1995b}.
Gallavotti's focus is on small fluctuations from zero forcing in a
steady state system whereas ours is on small fluctuations of select
observables from equilibrium.  Critical to his results is the
assumption that the dynamics of the system satisfy the chaotic
hypothesis, from which the fluctuation theorem may be inferred.  By
contrast, the validity of our approach rests solely on the
differentiability of the dynamical free energy, as defined in Eq.\
(\ref{eqn:DFE}) below.


\section{Onsager Linear Relations}
\label{sec:OLR}

In this section, the Onsager linear relations will be derived for the
short-time, near-equilibrium regime.  For a system with $n$ degrees of
freedom, let $X_n$ denote the set of microstates along with its
associated topology.  If, say, $x_0$ is the initial microstate of the
system at time zero, then the state at time $t$ is denoted by
$\Phi_{n,t}(x_0)$, where $\Phi_{n,t}$ is a Borel measurable
transformation on $X_n$.  For simplicity we shall restrict attention
to cases for which the set $\set{\Phi_{n,t}}_{t\in\Rset}$ forms a
group, though many results generalize without this assumption.  By
$G_n$ we denote the macroscopic observable of interest, which is
assumed to be a Borel measurable function from $X_n$ to $\Rset^d$.
Given an initial microstate $x_0$, then, $G_n(x_0) = a_0$ is the
initial value of all macroscopic variables under consideration.  As
only $a_0$, not $x_0$, is given, an \textit{a priori} probability
measure $P_n$ over the Borel subsets of $X_n$ is assumed to be valid
and invariant under $\Phi_{n,t}$.  This last point introduces a
statistical description of the system.


\subsection{Large Deviation Results}

The sequence of macroscopic observables $\seq{G_n}_{n\in\Nset}$ is
assumed to scale as $v_n$, where $\seq{v_n}_{n\in\Nset}$ is positive
and unbounded.  (For example, if $G_n$ is a sample mean then $v_n=n$.)
For $G_{n,t} \equiv G_{n}\circ\Phi_{n,t}$ and $G_{n,0} = G_n$, we
defined in \cite{LaCour_and_Schieve2002} a quantity $\Psi_t$, called
the dynamical free energy, where
\begin{equation}
\label{eqn:DFE}
\Psi_{t}(\lambda,\nu) \equiv \lim_{n\to\infty}
v_n^{-1}\log\avg{\e^{v_n[\lambda G_{n,0}+\nu G_{n,t}]}},
\end{equation}
Here, the values of $G_{n,0}$ and $G_{n,t}$ are considered to be
column vectors while the conjugate macrostates, $\lambda$ and $\nu$
are taken as row vectors; the symbol $\avg{\cdot}$ denotes expectation
with respect to the \textit{a priori} measure $P_n$.

We suppose that $\Psi_{t}$ is everywhere well defined, finite, and
differentiable.  (The validity of this assumption depends, of course,
on the microscopic dynamics, as given by $\Phi_{n,t}$, the macroscopic
observables, $G_{n}$, and the scaling parameter $v_n$.)  It then
follows from the G\"{a}rtner-Ellis theorem \cite{Dembo_and_Zeitouni}
that the sequence $\seq{(G_{n,0},G_{n,t})}_{n\in\Nset}$ satisfies a
large deviation principle (LDP) whose rate function is given by the
Legendre transform of $\Psi_t$.

Using a theorem regarding conditional LDPs
\cite{LaCour_and_Schieve2002}, it further follows that $G_{n,t}$, when
conditioned on a value $a_0$ of $G_{n,0}$, converges in probability as
$n\rightarrow\infty$ to the value $\nabla_{\nu}\Psi_t(\lambda_0,0)$,
where $\lambda_0$ satisfies $a_0 =
\nabla_{\lambda}\Psi_t(\lambda_0,0)$ \cite{LaCour_and_Schieve2002}.
The macroscopic map, $\psi_t$, associating an initial macrostate $a_0$
with a final macrostate $\psi_t(a_0)$ may then be defined as
\begin{equation}
\psi_t(a_0) \equiv \nabla_{\nu}\Psi_t(\lambda_0,0) =
\lim_{n\rightarrow\infty} \frac{\avg{G_{n,t}\,\e^{v_n\lambda_0
G_{n}}}}{\avg{\e^{v_n\lambda_0 G_{n}}}}.
\end{equation}
where $\lambda_0$ is uniquely defined by
\begin{equation}
a_0 = \nabla\Psi(\lambda_0) = \lim_{n\rightarrow\infty}
\frac{\avg{G_{n}\,\e^{v_n\lambda_0 G_{n}}}}{\avg{\e^{v_n\lambda_0
G_{n}}}}
\end{equation}
and $\Psi(\cdot) \equiv \Psi_t(\cdot,0)$ is the equilibrium free
energy.  We assume the Jacobian of $\nabla\Psi$ is nowhere zero, so
the mapping $\lambda_0 \mapsto a_0$ is invertible.  Large deviation
theory therefore implies that the most likely future macrostate is
given by a canonical expectation.  Note that for $a_0$ equal to the
equilibrium value $a_*$, where
\begin{equation}
a_* \equiv \lim_{n\rightarrow\infty}\avg{G_n} = \nabla\Psi(0),
\end{equation}
the corresponding conjugate macrostate is $\lambda_0 = 0$.


\subsection{Linearization}

The basic starting assumption is that the macroscopic variable
corresponds to the predicted macrostate $\psi_t(a_0)$.  Since a
one-to-one correspondence between $a_0$ and the conjugate macrostate
$\lambda_0$ holds, this macrostate may alternately be written as a
function $\alpha$ of $\lambda_0$ and $t$; thus, $\alpha(t,\lambda_0)
\equiv \psi_t(a_0)$.  Assuming this function to be analytic, a series
expansion is then made about equilibrium ($\lambda_0=0$) and the
initial time ($t=0$).  A linear law is thereby obtained which is valid
for short times and near-equilibrium initial conditions.  By equating
the equilibrium large deviation rate function with the thermodynamic
entropy, the flow and force will be identified, thereby establishing
the Onsager relation.  Throughout the derivation, no \textit{ad hoc}
mesoscopic time is introduced.  Finally, a expression for the Onsager
coefficients is derived in terms of a correlation matrix, which is
identified as a Green-Kubo formula, and the reciprocity relations then
follow from time reversal invariance.

Expanding $\alpha(t,\lambda_0)$ about $(0^{\pm},0)$, where $0^{\pm}$
indicates a right/left-sided time derivative evaluated at zero, we
find, to second order,
\begin{equation}
\begin{split}
\alpha(t,\lambda_0) &\approx \left.\alpha\right|_{(0^{\pm},0)} +
\left.(t\partial_t + \lambda_0\nabla)\alpha\right|_{(0^{\pm},0)} + \frac{1}{2}\left.(t\partial_t +
\lambda_0\nabla)^2\alpha\right|_{(0^{\pm},0)} \\
&= a_* + \left.(\lambda_0\nabla)\alpha\right|_{(0^{\pm},0)} +
\left.(t\partial_t\lambda_0\nabla)\alpha\right|_{(0^{\pm},0)} +
\frac{1}{2}\left.(\lambda_0\nabla)^2\alpha\right|_{(0^{\pm},0)}.
\end{split}
\end{equation}
The second line in the above display results from the fact that
$\partial_t\alpha(0^{\pm},0)$ and $\partial_t^2\alpha(0^{\pm},0)$ are
both zero, since $\alpha(t,0) = a_*$.  (Recall that $P_n$ is invariant
under $\Phi_{n,t}$.)  As all terms but the third are independent of
$t$ (to second order), the partial time derivative of
$\alpha(t,\lambda_0)$ is
\begin{equation}
\partial_t\alpha(t,\lambda_0) \approx
\left.(\lambda_0\,\partial_t\nabla)\alpha\right|_{(0^{\pm},0)}.
\end{equation}

The left-hand side is identified with the thermodynamic flow, $J(t)$.
(This follows from the identification of $\alpha(t,\lambda_0)$ as the
macroscopic variable.)  To identify $L$ and $X$ on the right hand
side, recall that $X = \transpose{\nabla}S$ by definition.  If we
postulate that $S = -I$, where $I$ is the equilibrium rate function,
i.e., the Legendre transform of $\Psi$, then $a_0 =
\nabla\Psi(\lambda_0)$ implies that $\lambda_0 =
\transpose{\nabla}I(a_0) = -\transpose{\nabla}S(a_0) = -X(a_0)$.
Substituting, we find
\begin{equation}
J(t) \approx
\left.(-X(a_0)\,\partial_t\nabla)\alpha\right|_{(0^{\pm},0)},
\label{eqn:flow-force}
\end{equation}
or, in explicit component form,
\begin{equation}
J_i(t) \approx \sum_{j} -\left.\frac{\partial^2\alpha_i}{\partial t
\partial\lambda_j}\right|_{(0^{\pm},0)} X_j(a_0).
\end{equation}
Evidently, the Onsager matrix is given by
\begin{equation}
L^{(\pm)} = \left.
\transpose{(-\partial_t\nabla\transpose{\alpha})}\right|_{(0^{\pm},0)}.
\label{eqn:Onsager_matrix}
\end{equation}

For a single (i.e., scalar) observable, a positive $\lambda_0$
corresponds to an initial macrostate above equilibrium, while a
negative $\lambda_0$ corresponds to one below equilibrium.  Therefore,
assuming $L^{(+)}$ is positive, the initial tendency of $\alpha$ is to
move towards equilibrium.  (For large $n$ this means that the actual
observable, $G_{n,t}$, tends towards equilibrium with probability near
one.) If $\alpha$ is symmetric in time, due for example to time
reversal invariance (see \cite{LaCour_and_Schieve2002}), then the same
holds true in the reverse time direction as well.


\subsection{Semigroups and Equilibration}

The initial tendency towards equilibrium does not necessarily imply a
long-time approach to equilibrium unless the family
$\set{\psi_t}_{t\in\Rset}$ of macroscopic maps forms a semigroup.  As
an aside, suppose the family of macroscopic maps does in fact form a
semigroup, i.e., that $\psi_{t+s}(a_0) = \psi_s(\psi_t(a_0))$ for $s,t
\ge 0$ (or $s,t \le 0$), and suppose further that $\psi_t(a_0)$ is
near the equilibrium value, $a_*$.  The linear approximation then
leads to an exponential time evolution for the macrostate, since
\begin{equation}
\begin{split}
\dot{\psi}_t(a_0)&=\lim_{s\to0}\frac{\psi_{t+s}(a_0)-\psi_t(a_0)}{s}\\
&= \lim_{s\to0}\frac{\psi_{s}(\psi_t(a_0))-\psi_t(a_0)}{s} \approx L
\, \nabla S(\psi_t(a_0)) \\
&\approx L\left\{ \nabla S(a_*) + \nabla\transpose{\nabla}S(a_*) \,
[\psi_t(a_0)-a_*] \right\} \\
&= L \, \nabla\transpose{\nabla}S(a_*) \, [\psi_t(a_0)-a_*]
\end{split}
\end{equation}
implies
\begin{equation}
\psi_t(a_0) \approx a_* + (a_0 - a_*)
\exp[tL\nabla\transpose{\nabla}S(a_*)].
\end{equation}
The essence of the time-averaging approach used by Casimir, Mori,
Oono, and others is to suppose that there is a mesoscopic time $\tau$
large enough so that $\psi_{t+\tau}(a_0) \approx
\psi_{\tau}(\psi_t(a_0))$ yet small enough so that $\dot{\psi}_t(a_0)
\approx [\psi_{t+\tau}(a_0)-\psi_t(a_0)]/\tau$.  If this can be shown,
then the usual exponential macroscopic law, as derived above, follows.
The precise conditions for the validity of this approximation will not
be pursued here further, however.


\section{Green-Kubo Formula}
\label{sec:GKF}

Having obtained an explicit expression for the Onsager matrix in terms
of $\alpha$, the dynamical free energy may now be used to derive a
Green-Kubo formula, giving $L^{(\pm)}$ in terms of the rate of change
of the asymptotic covariance between $G_n$ and $G_{n,t}$.  We begin by
observing that, since $\alpha(t,\lambda) =
\nabla_{\nu}\Psi_t(\lambda,0)$,
\begin{equation*}
\left.(\nabla\transpose{\alpha})\right|_{(t,0)} = \left. \left[
\nabla_{\lambda} \lim_{n\to\infty} \avg{\transpose{G_{n,t}}}_{\lambda}
\right] \right|_{\lambda=0},
\end{equation*}
where
\begin{equation}
\avg{G_{n,t}}_{\lambda} \equiv \frac{\avg{G_{n,t}\,\e^{v_n\lambda
G_{n}}}}{\avg{\e^{v_n\lambda G_{n}}}}.
\end{equation}
Convexity of the free energy allows us to take the derivative within
the limit, so
\begin{equation*}
\begin{split}
\left.(\nabla\transpose{\alpha})\right|_{(t,0)} &= \lim_{n\to\infty}
v_n \!\left.\left[ \avg{G_{n}\transpose{G_{n,t}}}_{\lambda} \!\!-
\avg{G_{n}}_{\lambda} \avg{\transpose{G_{n,t}}}_{\lambda}
\right]\right|_{\lambda\!=0} \\
&= \lim_{n\to\infty} v_n \left[ \avg{G_{n}\transpose{G_{n,t}}} -
\avg{G_{n}} \avg{\transpose{G_{n,t}}} \right].
\end{split}
\end{equation*}
From the above result and Eq.\ (\ref{eqn:Onsager_matrix}), the
Green-Kubo relation now follows:
\begin{equation}
L^{(\pm)} = - \left.\frac{\d}{\d t} \lim_{n\to\infty} v_n
\left[\avg{G_{n,t}^{} \transpose{G_{n}}} -
\avg{G_{n,t}^{}}\avg{\transpose{G_{n}}}\right]\right|_{t=0^{\pm}}.
\label{eqn:Green-Kubo}
\end{equation}

This result is equivalent to that of Oono in Eq.~(\ref{eqn:Oono}),
assuming $\avg{G_n} = 0$ and $G_{n,t}$ is differentiable almost
everywhere.  Furthermore, since the \textit{a priori}\ measure is
invariant, the Onsager matrix may be written in the following, more
explicit form.
\begin{equation}
L^{(\pm)} = -\lim_{t\to0^{\pm}} \lim_{n\to\infty} \frac{v_n}{t} \left[
\avg{G_{n,t} \transpose{G_{n}}} - \avg{G_{n}} \avg{\transpose{G_{n}}}
\right].
\label{eqn:simple_Green-Kubo}
\end{equation}


\subsection{Time Reversal Invariance}

Using Eq.~(\ref{eqn:Green-Kubo}) or (\ref{eqn:simple_Green-Kubo}),
the Onsager reciprocal relations can be shown to follow from time
reversal invariance and measure preservation.  Specifically, if
$\Phi_{n,t}$ is time reversal invariant under some map\ $R_{n}$ and
both $P_n$ and $G_{n}$ are invariant under $R_{n}$, then $L^{(\pm)}$
is symmetric.  To prove this, it suffices to show that
$\avg{G_{n,t}^{}\transpose{G_{n}}} =
\avg{G_{n}^{}\transpose{G_{n,t}}}$.  First note that
\begin{equation}
\begin{split}
\avg{G_{n,t}^{}\transpose{G_{n}}}
&= \avg{(G_{n}^{}\circ R_{n}^{}\circ\Phi_{n,t}^{})\,\transpose{G_{n}}} \\
&= \avg{(G_{n}^{}\circ\Phi_{n,-t}^{}\circ
   R_{n}^{})(\transpose{G_{n}}\circ R_{n}^{})} \\
&= \avg{G_{n,-t}^{}\transpose{G_{n}}}.
\end{split}
\end{equation}
Using this result and the fact that $P_n$ is invariant, we find
\begin{equation}
\begin{split}
\avg{G_{n,-t}^{}\transpose{G_{n}}} &=
  \avg{(G_{n}^{}\circ\Phi_{n,-t}^{})\,\transpose{G_{n}}} \\
&= \avg{G_{n}^{}\,\transpose{(G_{n}^{}\circ\Phi_{n,t}^{})}} =
\avg{G_{n}^{}\transpose{G_{n,t}}}.
\end{split}
\end{equation}
Combining the above two equations then gives the desired result that
$L^{(\pm)}$ is symmetric.  Note that time reversal invariance is not a
necessary condition for the reciprocity relations to hold, as shown in
\cite{Gabrielli1996}.

Time reversal invariance, as defined above, is known to imply time
symmetry for the predicted macroscopic time evolution; i.e.,
$\alpha(-t,\cdot) = \alpha(t,\cdot)$.  It is therefore not surprising
that $L^{(+)}$ and $L^{(-)}$ should reflect that symmetry.  Indeed, it
is clear that $L^{(-)} = -L^{(+)}$ since
\begin{equation}
\begin{split}
L^{(-)} &= -\lim_{t\to0^-} \lim_{n\to\infty} \frac{v_n}{t} \left[
  \avg{G_{n,t} \transpose{G_{n}}} - \avg{G_{n}}
  \avg{\transpose{G_{n}}} \right] \\
&= -\lim_{t\to0^+} \lim_{n\to\infty} \frac{v_n}{-t} \left[
  \avg{G_{n,-t} \transpose{G_{n}}} - \avg{G_{n}}
  \avg{\transpose{G_{n}}} \right] \\
&= \lim_{t\to0^+} \lim_{n\to\infty} \frac{v_n}{t} \left[ \avg{G_{n,t}
   \transpose{G_{n}}} - \avg{G_{n}} \avg{\transpose{G_{n}}} \right] \\
&= -L^{(+)}.
\end{split}
\end{equation}
This result may be viewed as a statement of the Onsager-Machlup
principle.


\subsection{Positive Semidefinite Matrix}

The forward Onsager matrix, $L^{(+)}$, may be shown to be positive
semidefinite under more general conditions, using only the assumption
that the \textit{a priori}\ measure is invariant.  To see this,
observe that for any given row vector $X$
\begin{equation}
X L^{(+)} \transpose{X} = -\lim_{t\to0^{\pm}} \lim_{n\to\infty}
\frac{v_n}{t} \left[ \avg{XG_{n,t} \transpose{G_{n}}\transpose{X}} -
\avg{XG_{n}} \avg{\transpose{G_{n}}\transpose{X}} \right].
\end{equation}
Note that $X G_{n,t}^{}$ and $\transpose{G_{n}}\transpose{X} =
\transpose{(X G_{n}^{})} = X G_{n}^{}$ are both scalar-valued random
variables.  The Cauchy-Schwarz inequality therefore implies that
\begin{equation}
\begin{split}
\avg{X G_{n,t}^{} X G_{n}^{}} &\le \avg{(X G_{n,t}^{})^2}^{1/2}
\avg{(X G_{n}^{})^2}^{1/2} = \avg{(X G_{n}^{})^2},
\end{split}
\end{equation}
where the final equality is due to the invariance of the \textit{a
priori} measure.

Since $\avg{X G_{n,t}^{} \transpose{G_{n}} \transpose{X}} = \avg{X
G_{n,t}^{} X G_{n}^{}}$ and $\avg{X G_{n}} \avg{\transpose{G_{n}}
\transpose{X}}= \avg{(X G_{n}^{})^2}$, we conclude that $X L^{(+)}
\transpose{X} \ge 0$ for any $X$, so $L^{(+)}$ is positive
semidefinite.


\subsection{Entropy Production Rate}

We end this section with a brief discussion of the internal entropy
production rate, $\sigma(t,\lambda_0) \equiv
\dot{S}(\alpha(t,\lambda_0))$, which is an important measure of
irreversibility in nonequilibrium thermodynamics.  Its nonnegativity
is traditionally derived from an extrapolation of the equilibrium
second law to the nonequilibrium regime \cite{deGroot_and_Mazur}; here
we derive it directly.

From Eqs.~(\ref{eqn:entropy_rate}) and (\ref{eqn:Onsager_relation}),
together with Eqs.~(\ref{eqn:flow-force}) and
(\ref{eqn:Onsager_matrix}), the initial entropy production rate is
\begin{equation}
\begin{split}
\sigma(0^{\pm},\lambda_0) &\equiv \lim_{t\rightarrow0^{\pm}}
[S(\alpha(t,\lambda_0))-S(\alpha(0,\lambda_0))]/t \\
&= X(a_0)J(0^{\pm}) \approx \lambda_0 L^{(\pm)} \transpose{\lambda_0}.
\end{split}
\end{equation}
For a system which is time reversal invariant, $L^{(+)}$, is positive
semidefinite and $L^{(-)} = -L^{(+)}$ is negative semidefinite, so the
nonequilibrium second law follows:
\begin{equation}
\sigma(0^+,\lambda_0) \ge 0, \quad \sigma(0^-,\lambda_0) \le 0.
\end{equation}
Note that the above inequalities are valid only for early times near
equilibrium.

The first inequality, $\sigma(0^+,\lambda_0) \ge 0$, tells us that the
initial tendency of the system is to evolve to a macrostate of higher
entropy.  Since $S = -I$, this means that the macrostate initially
tends to move closer to the minimum of $I$, i.e., the equilibrium
value $a_*$.  Similarly, the second inequality, $\sigma(0^-,\lambda_0)
\le 0$, means that nonequilibrium states tend to arise from states
closer to equilibrium in the near past.  These interpretations of
macroscopic fluctuations are entirely consistent with those of
Boltzmann's $H$-theorem as given long ago by P. and T. Ehrenfest
\cite{Ehrenfests}, provided $H$ is defined as a function on the phase
space rather than as a functional of the probability density.


\section{Ideal Gas Example}
\label{sec:IGE}

As an illustration of the above results, we consider the case of an
ideal gas in a rigid box with an imaginary partition in the middle.
The particles evolve freely and interact only with the walls of the
container via specular reflection.  Such a system is represented by a
Knundsen gas, for which the mean free path is much larger than the
linear dimension of the container, so particle collisions may be
ignored \cite{Kogan}.  The observable is taken to be simply the
fraction of particles on one side of the imaginary partition.  The
\textit{a priori} spatial distribution of the particles is uniform,
whereas that of the velocity is direction independent and given by the
symmetric probability density function $\phi_{\sigma}$.

In \cite{LaCour_and_Schieve2000} it was shown that the macroscopic map
for this observable is
\begin{equation}
\psi_t(a_0) = a_0 \sum_{n=-\infty}^{\infty}I_n(t) + (1-a_0)
\sum_{n=-\infty}^{\infty}\bar{I}_n(t),
\end{equation}
where
\begin{subequations}
\begin{equation}
I_{n}(t) \equiv 2 \int_{-\infty}^{\infty}\int_{0}^{1/2}
\!1_{[0,1/2]}(\xi\!-\!\eta t\!-\!n) \, \phi_{\sigma}(\eta) \, \d\xi
\d\eta
\end{equation}
if $n$ is even and, if $n$ is odd,
\begin{equation}
I_{n}(t) \equiv 2 \int_{-\infty}^{\infty}\int_{0}^{1/2}
\!1_{[0,1/2]}(1\!-\!\xi\!+\!\eta t\!+\!n) \, \phi_{\sigma}(\eta) \,
\d\xi \d\eta.
\end{equation}
\end{subequations}
($1_{[0,1/2]}$ is the indicator function on the interval $[0,1/2]$.)
For $\bar{I}_{n}(t)$, similar definitions hold with, however,
$1_{[0,1/2]}$ replaced by $1_{[1/2,1]}$.

For $t$ positive and small we find that
\begin{equation}
\psi_t(a_0) = a_0\left[I_0(t)+I_{-1}(t)\right] +
(1-a_0)\left[\bar{I}_0(t)+\bar{I}_{-1}(t)\right] + \cdots,
\end{equation}
which, since $\lambda_0 = \log[a_0/(1-a_0)]$, may be written
\begin{equation}
\alpha(t,\lambda_0) =
\frac{\e^{\lambda_0}\left[I_0(t)+I_{-1}(t)\right] +
\left[\bar{I}_0(t)+\bar{I}_{-1}(t)\right]}{1+\e^{\lambda_0}} + \cdots.
\end{equation}
Expanding $\alpha$ about $(0^{\pm},\lambda_0)$ gives
\begin{equation}
\alpha(t,\lambda_0) = \frac{\e^{\lambda_0} -
2(\e^{\lambda_0}-1)L\abs{t}}{1+\e^{\lambda_0}} + \cdots,
\end{equation}
where
\begin{equation}
L \equiv \int_{0}^{\infty} \eta \, \phi_{\sigma}(\eta) \, \d\eta,
\end{equation}
so expanding about $\lambda_0 = 0$ finally gives
\begin{equation}
\alpha(t,\lambda_0) = \frac{1}{2} + \left(\frac{1}{4} -
L\abs{t}\right)\lambda_0 + \cdots.
\end{equation}
By inspection we identify $L$ with the forward Onsager coefficient
$L^{(+)}$.  Note that, from its definition, $L$ is strictly positive
since $\phi_{\sigma}$ was assumed to be symmetric.

For an \textit{a priori} Maxwell-Boltzmann velocity distribution,
\begin{equation}
\phi_{\sigma}(\eta) = (2\pi\sigma^2)^{-1/2}\,\e^{-\eta^2/(2\sigma^2)},
\end{equation}
the Onsager coefficient is
\begin{equation}
L = \frac{\sigma}{\sqrt{2\pi}}=\frac{1}{\ell}\sqrt{\frac{kT}{2\pi m}},
\end{equation}
where $T$ is the absolute temperature, $m$ is the particle mass, and
$\ell$ is the length of the box.  The initial entropy production rate
is therefore
\begin{equation}
\sigma(0^{\pm},\lambda_0) = \pm\frac{1}{\ell}\sqrt{\frac{kT}{2\pi m}}
\left[\log\left(\frac{a_0}{1-a_0}\right)\right]^2.
\label{eqn:idealgas_entropy_production}
\end{equation}
Similarly, for an \textit{a priori} velocity distribution which is
uniform over the interval $[-\Delta,\Delta]$, we find $L = \Delta/4$.
The equilibrium entropy in both cases is
\begin{equation*}
S(a_0) = - \left[ a_0\log a_0 + (1-a_0)\log(1-a_0) + \log 2 \right],
\end{equation*}
where $a_0$ is the initial fraction of particles in the interval
$[0,\ell/2]$ (i.e., the left side).  Note that this is the entropy
derived from the large deviation rate function for this particular
observable and is not to be confused with the standard Sackur-Tetrode
equation of an ideal gas.

\begin{figure}
\centerline{\scalebox{.5}{\includegraphics{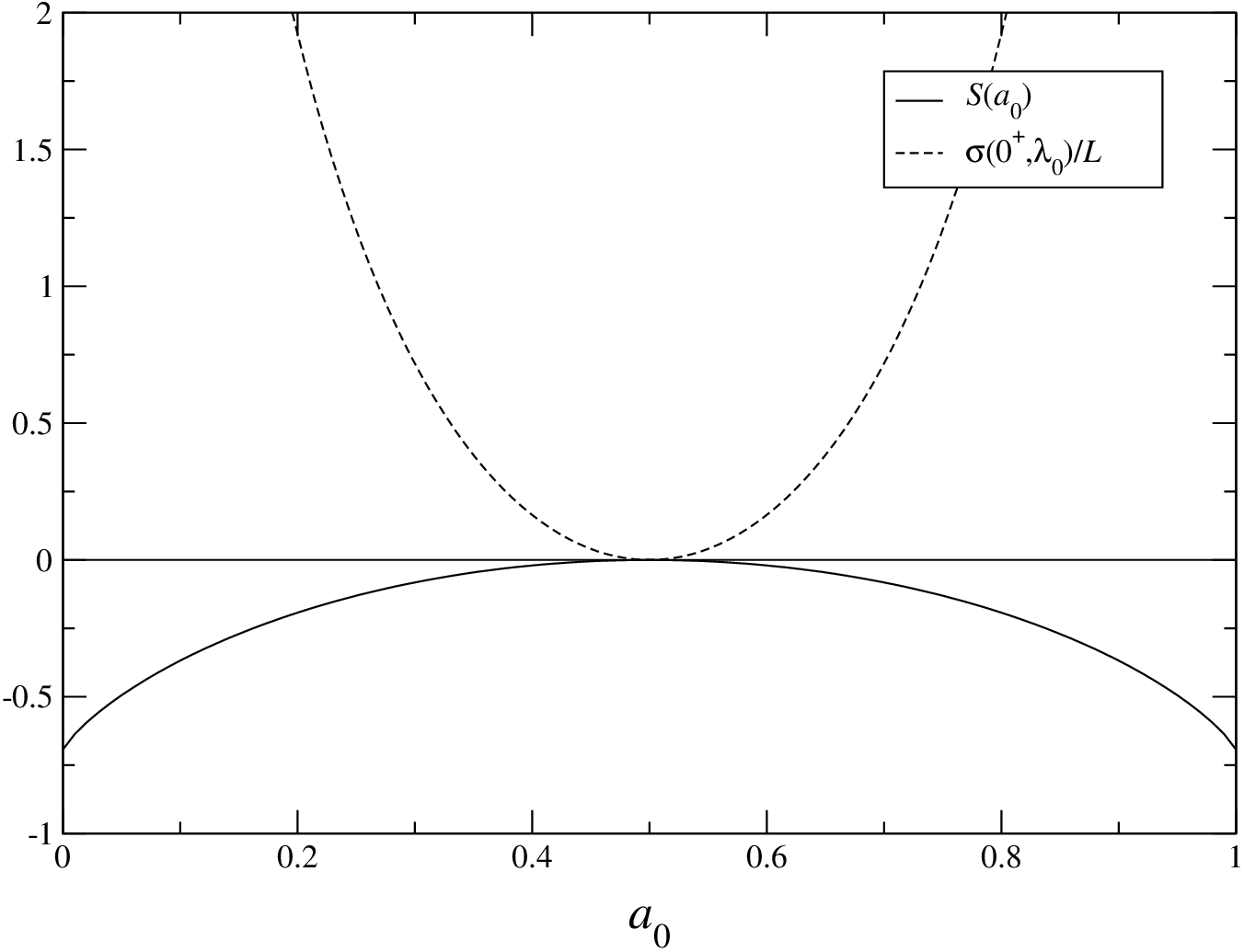}}}
\caption[Equilibrium entropy and entropy rate of an ideal gas]{Plot of
the equilibrium entropy, \protect$S(a_0)$, (solid curve) and rescaled
initial entropy production rate, \protect$\sigma(0^+,\lambda_0)/L$,
(dashed curve) versus the initial fractional occupation,
\protect$a_0$, of ideal gas particles occupying one side of a rigid
box.}
\label{fig:entropy}
\end{figure}

In Fig.~\ref{fig:entropy} are plotted the equilibrium entropy,
$S(a_0)$, and the initial forward entropy production rate,
$\sigma(0^+,\lambda_0)$.  (The latter has been rescaled to remove the
incidental factor of $L$.)  These graphs correspond to an ideal gas in
a rigid box (or cylinder) with an arbitrary symmetric velocity
distribution and independent uniform spatial distribution.  The
symmetry of the graphs is due to the symmetry of the imaginary box
partition.  The peak at $a_0=1/2$ in $S$ corresponds to the most
likely \textit{a priori} configuration.  The entropy rate takes its
minimum value of zero at this location, indicating that there is no
tendency to deviate from an initial equilibrium macrostate.  For
$a_0\neq1/2$, the entropy production rate is strictly positive,
indicating an initial tendency to approach equilibrium.  At the
extremal points, $a_0=0\text{ or }1$, the entropy production rate
diverges; however, Eq.~(\ref{eqn:idealgas_entropy_production}) is
strictly valid only near equilibrium.

\begin{figure}
\centerline{\scalebox{.5}{\includegraphics{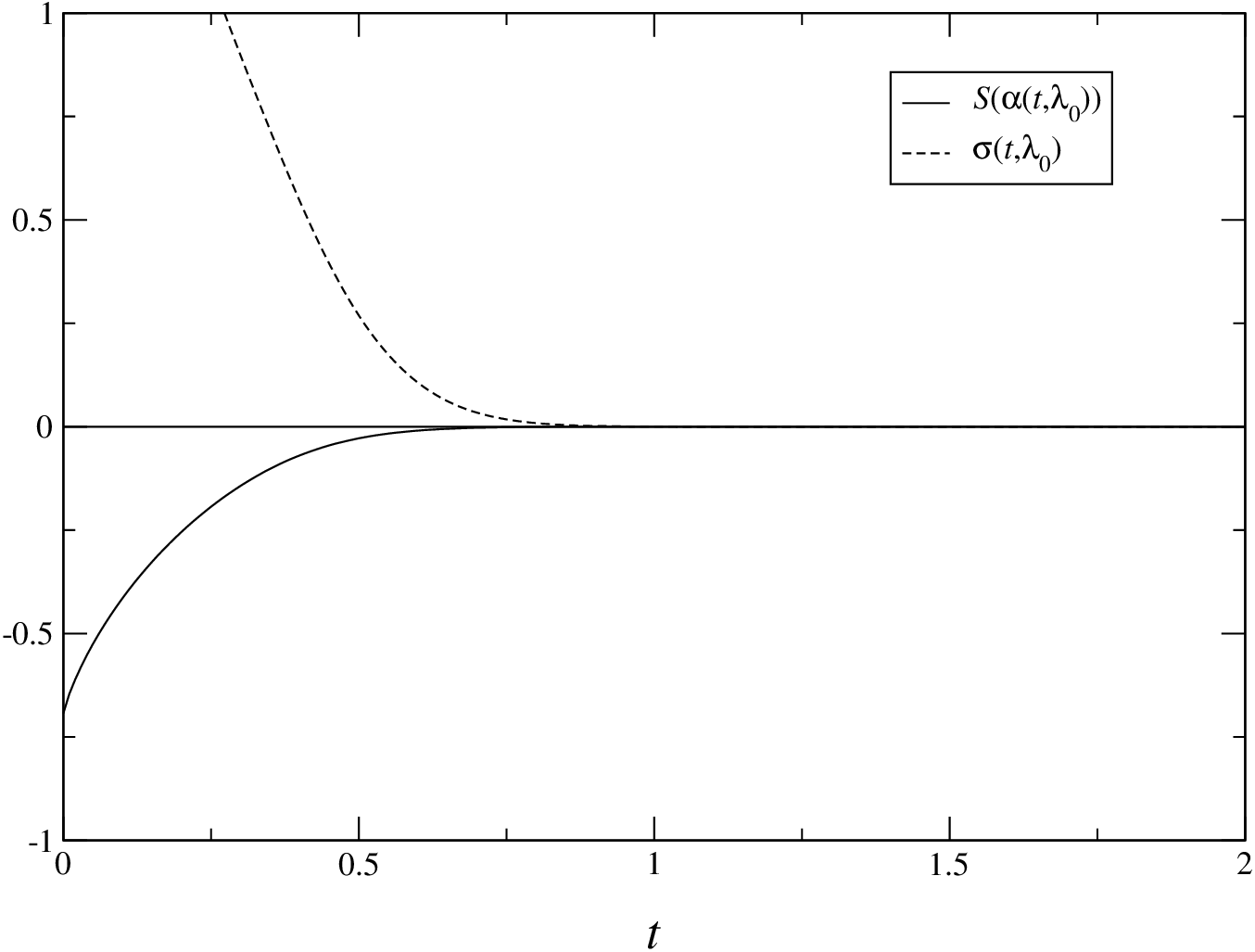}}}
\caption[Dynamic entropy of an ideal gas with a Maxwellian
distribution]{Plot of the dynamic entropy,
\protect$S(\alpha(t,\lambda_0))$, (solid curve) and entropy production
rate, \protect$\sigma(t,\lambda_0)$, (dashed curve) versus time for an
ideal gas with an \textit{a priori} Maxwellian velocity distribution
of unit variance.  The initial macrostate is \protect$a_0=1 \;
(\lambda_0=\infty)$.}
\label{fig:maxgas_entropy}
\end{figure}

In Fig.~\ref{fig:maxgas_entropy} we have plotted the dynamic entropy,
$S(\alpha(t,\lambda_0))$, and entropy production rate,
$\sigma(t,\lambda_0)$, (computed numerically from the former) versus
time for an ideal gas with $a_0=1$ and an \textit{a priori} Maxwellian
velocity distribution.  (The picture is qualitatively the same for
other values of $a_0$.)  The approach to equilibrium is monotonic, as
indicated by the fact that $\sigma(t,\lambda_0) \ge 0$, in accordance
with the usual extrapolation of the second law beyond equilibrium.

\begin{figure}
\centerline{\scalebox{.5}{\includegraphics{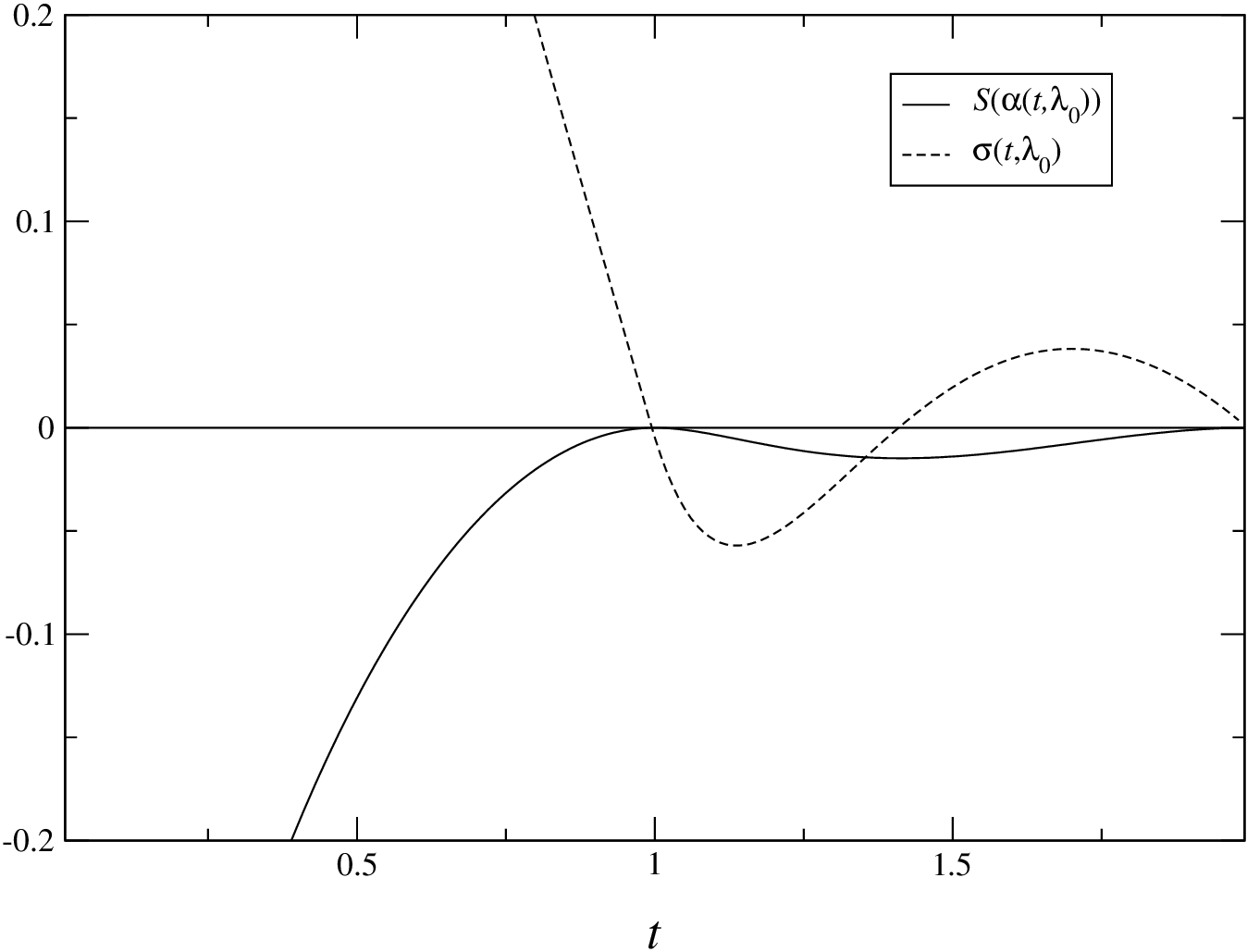}}}
\caption[Dynamic entropy of an ideal gas with a uniform
distribution]{Plot of the dynamic entropy,
\protect$S(\alpha(t,\lambda_0))$, (solid curve) and entropy production
rate, \protect$\sigma(t,\lambda_0)$, (dashed curve) versus time for an
ideal gas with an \textit{a priori} uniform velocity distribution with
support on \protect$[-1,1]$.  The initial macrostate is \protect$a_0=1
\; (\lambda_0=\infty)$.}
\label{fig:unifgas_entropy}
\end{figure}

The following picture, Fig.~\ref{fig:unifgas_entropy}, shows the same
quantities but with an \textit{a priori}\ velocity distribution which
is uniform on the interval $[-1,1]$.  After an initial monotonic
approach to equilibrium, the system exhibits ``anti-thermodynamic''
behavior from $t=1$ to about $t=1.4$.  For larger values of $t$ the
entropy productions rate continues to oscillate about zero, in step
with the oscillations of $\alpha(t,\lambda_0)$ about $1/2$, but
converges to zero as $t\to\infty$.  The oscillations may be
interpreted as an indication of phase mixing, wherein the phase space
becomes filibrated due solely to the dispersion of particles traveling
at different, yet constant, velocities.  An important point is that,
despite their anomalous behavior, these graphs are consistent with a
positive Onsager coefficient, since the antithermodynamic behavior
occurs for $t$ far from zero.

A comparison of Figs.~\ref{fig:maxgas_entropy}
and~\ref{fig:unifgas_entropy} suggests that a system of nonideal gas
particles, e.g.\ rigid spheres, should give rise to a dynamic entropy
which is monotonic beyond an initial equilibration phase in which, in
accordance with the Boltzmann equation, the particles should attain an
approximately Maxwellian velocity distribution.  Given the very short
time scale over which this local equilibration is expected to occur,
it is not surprising that antithermodynamic behavior such as we have
described is not normally observed.

\section{Discussion}
\label{sec:Discussion}

We have considered the time evolution of a given set of macroscopic
variables in a microscopically deterministic, isolated system.  At
equilibrium, these variables attain their most probable values but
over time will exhibit fluctuations away from equilibrium.  We have
shown that for small fluctuations the most probable evolution of these
fluctuations in the near future (and near past) is governed by the
Onsager principle, which gives a linear relationship between the flows
and thermodynamic forces.  The former was interpreted as the time rate
of change of the most probable macrostate, while the latter was found
to correspond to the Legendre-transform conjugate of the initial,
nonequilibrium macrostate.  This result was obtained using
well-established techniques from large deviation theory and is
independent of any \textit{ad hoc} mesoscopic time scale.  An explicit
expression for the forward and reverse Onsager matrices in terms of
the covariance matrix of the macroscopic variables then followed.

The forward Onsager matrix was found to be positive semidefinite as a
consequence of the assumed invariance of the \textit{a priori}
probability measure.  The Onsager reciprocity relations were verified
in the case of time reversal invarinace and found to imply that the
forward and reverse Onsager matrices differ only by a sign.  Together,
these last two results imply that the second law of thermodynamics, as
regards fluctuations, is valid in the aforementioned regime for both
the forward and reverse time directions.  Thus, given an initial
nonequilibrium macrostate, a system's initial tendency is to approach
equilibrium (in the forward time direction) or to have arisen from a
state closer to equilibrium (in the reverse time direction).  This
need not imply long-time approach to equilibrium, however, unless, for
example, the macroscopic dynamics form a semigroup.  At much later or
earlier times, antithermodynamic behavior is also possible, as
illustrated by the example considered of an ideal gas with a uniform
velocity distribution.


\section*{Acknowledgments}

This work was supported in part by the Engineering Research Program of
the Office of Basic Energy Sciences at the U.S.\ Department of Energy,
Grant No.\ DE-FG0394ER14465.  One of us (B.L.) has also received
partial funding from Applied Research Laboratories of the University
of Texas at Austin, Independent Research and Development Grant No.\
926.



\end{document}